\documentclass[aps,reprint,superscriptaddress,amsmath,amssymb,longbibliography]{revtex4-1}
\usepackage{titlesec}

\usepackage{graphicx}
\usepackage{amsmath}
\usepackage{bm}

\usepackage{color}
\definecolor{CLBlue}{rgb}{0, .3, .6}

\begin{document}

%\preprint{}

\title{Minimax entropy: The statistical physics of optimal models}
%\title{Minimax entropy: The statistical physics of useful models}


\author{David P. Carcamo}
\affiliation{Department of Physics, Yale University, New Haven, CT, USA}
\affiliation{Quantitative Biology Institute, Yale University, New Haven, CT, USA}
\author{Nicholas J. Weaver}
\affiliation{Department of Physics, Yale University, New Haven, CT, USA}
\affiliation{Quantitative Biology Institute, Yale University, New Haven, CT, USA}
\author{Purushottam D. Dixit}
\affiliation{Department of Biomedical Engineering, Yale University, New Haven, CT, USA}
\affiliation{Systems Biology Institute, Yale University, West Haven, CT, USA}
\author{Christopher W. Lynn}
\email{Corresponding author: christopher.lynn@yale.edu}
\affiliation{Department of Physics, Yale University, New Haven, CT, USA}
\affiliation{Quantitative Biology Institute, Yale University, New Haven, CT, USA}
\affiliation{Wu Tsai Institute, Yale University, New Haven, CT, USA}

%\thanks{}


\date{\today}

\begin{abstract}

When constructing models of the world, we aim for optimal compressions: models that include as few details as possible while remaining as accurate as possible. But which details---or features measured in data---should we choose to include in a model? Here, using the minimum description length principle, we show that the optimal features are the ones that produce the maximum entropy model with minimum entropy, thus yielding a minimax entropy principle. We review applications, which range from machine learning to optimal models of biological networks. Naive implementations, however, are limited to systems with small numbers of states and features. We therefore require new theoretical insights and computational techniques to construct optimal compressions of high-dimensional datasets arising in large-scale experiments.

\end{abstract}

%\pacs{asdfasdf}
%\keywords{keywords}

\maketitle

\section{Introduction}

Statistical models are inherently lossy compressions. When confronted with experimental data, one does not include every available feature. In practice, with finite samples, including too many features leads to overfitting. In theory, even without sampling errors, one should discard features that do not improve their description of a system. In this way, all models are wrong, but which are useful?

\begin{figure}[t]
    \centering
    \includegraphics[width=\linewidth]{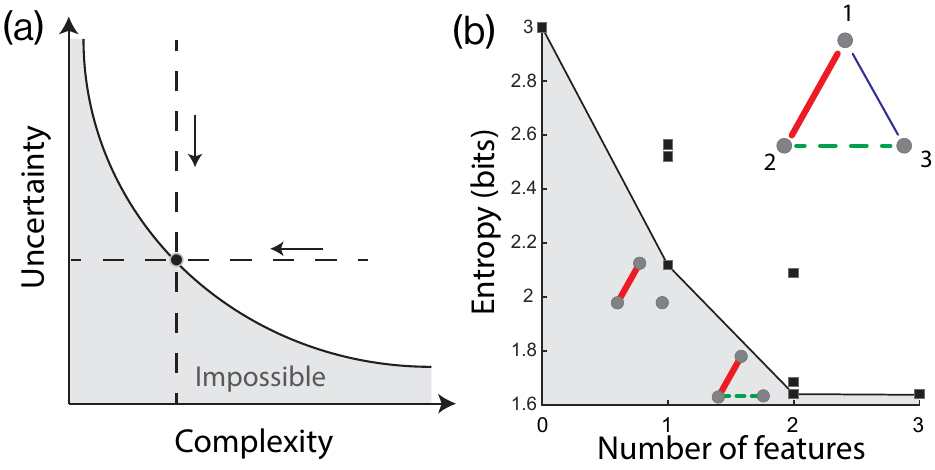}
    \caption{Useful models as optimal compressions. (a) Illustration of uncertainty versus complexity. Optimal models minimize uncertainty for a given complexity (vertical line) or, equivalently, minimize complexity for a given uncertainty (horizontal line). (b) For the three-spin Ising model in Eq.~(\ref{eq_Ising}), we plot the entropy $S(P)$ of maximum entropy models constructed to match different sets of pairwise correlations (square points). For each number of correlations, the optimal bound indicates the minimum entropy. Note that for maximum entropy models the entropy is equivalent to the description length $L(P)$ [Eqs.~(\ref{eq_LP1}-\ref{eq_LP2})].}
    \label{fig:figure1}
\end{figure}

Intuitively, useful models are those that strike an optimal balance between complexity and uncertainty. Given constraints on the complexity of a model---for example, the number of parameters or features included from data---one would like to provide the most accurate description of a system, thereby minimizing uncertainty [Fig.~\ref{fig:figure1}(a), vertical line]. Similarly, for a given level of uncertainty, one would like to remove as many details as possible, thus minimizing complexity [Fig.~\ref{fig:figure1}(a), horizontal line]. Importantly, for a given system, there is not a single useful model, but rather an entire spectrum. Where one lands on this spectrum depends on the relative value they assign to accuracy versus simplicity.

Information theory provides the mathematical framework to make these intuitions concrete \cite{Shannon-01, thomas_m_cover_elements_2006}. Each model---specifically, each distribution $P(x)$ over states $x$ of a system---defines an encoding of the system \cite{thomas_m_cover_elements_2006}. The length of this encoding, known as the description length, reflects the uncertainty of the modeler. Given a set of features measured in data and nothing else, the model with the minimum description length is the one with maximum entropy \cite{feder_maximum_1986}. Thus, the maximum entropy principle provides the most accurate mapping from experimental measurements to statistical models. However, it does not tell us which features (and therefore which model) one should select. For a given level of complexity---that is, under constraints on the number or type of features---which model minimizes our uncertainty?

%Yet across a range of possible features, which should we use to construct a model? Specifically, for a given level complexity---that is, under constraints on the number or type of features---which model minimizes our uncertainty?

%Yet across a range of possible features, which should we use to construct a model?

Here we show that the optimal features, which minimize description length, are the ones that produce the maximum entropy model with minimum entropy. This tells us that the best compressions identify the features in the data that maximally constrain our expectations, while remaining maximally random with regard to all unobserved features. Thus, starting only with the minimum description length (MDL) principle \cite{grunwald2007minimum, hansen2001model}, one can derive not only the maximum entropy principle, but also its generalization, the ``minimax entropy" principle.

Originally proposed in the context of machine learning \cite{zhu_minimax_1997, zhu_frame_1996}, the minimax entropy principle has recently emerged as a data-driven approach to constructing optimal models of complex living systems \cite{lynn2023exactly, lynn2023exact, carcamo_statistical_2024, Weaver-01}. These applications, however, are limited by our ability to solve statistical physics problems. We therefore require new theoretical and computational methods to study the high-dimensional data being generated in large-scale experiments.

Finally, even with the tools to solve the minimax entropy problem, there is no guarantee that a useful compression exists. This depends critically on the system of interest. Just as images of independent pixels cannot be compressed, some systems may not admit simplified models capable of making accurate predictions. Fortunately, converging evidence across physics, biology, and neuroscience provides hope for useful models that do not rely on all of the details.

The paper is organized as follows. In Sec.~\ref{sec_minimax}, we derive the minimax entropy principle as the solution to a compression problem. In Sec.~\ref{sec_opt}, we comment on alternative derivations, including minimizing the divergence from data and maximizing the information contained in experimental observations. In Sec.~\ref{sec_review}, we survey existing applications of the minimax entropy principle. In Sec.~\ref{sec_future}, we highlight key outstanding challenges, open questions, and future directions. Finally, in Sec.~\ref{sec_concl} we provide conclusions and outlook.

\section{Minimax entropy principle}
\label{sec_minimax}

\subsection{Maximizing entropy}

Consider a system of $N$ variables with collective states $\bm{x} = \{x_1,\hdots,x_N\}$. From experiments, we have access to features of the system, which can be represented as empirical averages $\langle f(\bm{x})\rangle_\text{exp}$, where $f(\bm{x})$ is an arbitrary function of the state $\bm{x}$. For example, we could build an empirical histogram by simply counting states, $P_\text{exp}(\bm{x}') = \langle \delta_{\bm{x},\bm{x}'}\rangle_\text{exp}$. Alternatively, we could measure the averages of individual variables $\langle x_i\rangle_\text{exp}$ or the correlations among pairs of variables $\langle x_ix_j\rangle_\text{exp}$, triplets $\langle x_ix_jx_k\rangle_\text{exp}$, or more. After measuring the averages of $M$ functions $\{f_\mu(\bm{x})\}$, where $\mu = 1,\hdots,M$, we would like to construct a model $P(\bm{x})$ that is consistent with these features, such that
\begin{equation}
\label{eq_f}
\langle f_\mu(\bm{x}) \rangle = \langle f_\mu(\bm{x}) \rangle_\text{exp} \quad \text{for all } \mu,
\end{equation}
where $\langle\cdot \rangle$ represents an average over $P(\bm{x})$. However, there are infinitely many distributions that match these features; how should we select one among them?

For every state $\bm{x}$, each model $P(\bm{x})$ defines a code word of length
\begin{equation}
\ell_P(\bm{x}) = -\log P(\bm{x}).
\end{equation}
The lower the probability $P(\bm{x})$, the more surprised the modeler is to see the state $\bm{x}$, and therefore the longer their encoding \cite{Shannon-01, thomas_m_cover_elements_2006}. To compute the total description length of the model, we must average over all hypothetical datasets that could have generated the observed features \cite{feder_maximum_1986}. Specifically, let $\{\bm{x}^{(t)}\}$ denote a dataset consisting of $T$ samples of the system state $\bm{x}^{(t)}$, where $t = 1,\hdots,T$. The set of all datasets consistent with the observed features is given by
\begin{equation}
\label{eq_chi}
\chi = \bigg\{ \{\bm{x}^{(t)}\} \, :\, \frac{1}{T}\sum_{t = 1}^T f_\mu (\bm{x}^{(t)}) = \langle f_\mu (\bm{x}) \rangle_\text{exp} \, \forall \,\mu \bigg\}.
\end{equation}
Averaging over these datasets, we arrive at the description length of the model,
\begin{equation}
\label{eq_L}
L(P) = \langle \ell_P(\bm{x})\rangle_\chi.
\end{equation}

To minimize the description length \cite{grunwald2007minimum, hansen2001model}, the model should be maximally unbiased; any source of order that is not justified by the measured features will lead to increases in code word lengths. Indeed, among all models consistent with the observed features $\mathcal{F} = \{f_\mu(\bm{x})\}$, one can show (in the limit of large $T$) that the description length $L(P)$ is minimized by the model with maximum entropy (Appendix \ref{app_MDL})
\begin{equation}
S(P) = -\sum_{\bm{x}} P(\bm{x}) \log P(\bm{x}).
\label{eq:ent}
\end{equation}
This maximum entropy model takes the form of a Boltzmann distribution \cite{Jaynes-01}
\begin{equation}
\label{eq_PF}
P_{\mathcal{F}}(\bm{x}) = \frac{1}{Z}\text{exp}\bigg[\sum_{\mu = 1}^M \lambda_\mu f_\mu(\bm{x})\bigg],
\end{equation}
where
\begin{equation}
Z = \sum_{\bm{x}} \text{exp}\bigg[\sum_\mu \lambda_{\mu = 1}^M f_\mu(\bm{x})\bigg]
\end{equation}
is the normalizing partition function (Appendix \ref{app_maxEnt}). The parameters $\lambda_\mu$ are Lagrange multipliers, which must be calculated so that the model matches the measured features [Eq.~(\ref{eq_f})]. This is the canonical inverse problem in statistical physics \cite{Jaynes-01, presse2013principles, nguyen2017inverse, zdeborova2016statistical, chayes1984inverse}.

The maximum entropy principle presents an exact path from measured features to statistical physics models \cite{Jaynes-01, presse2013principles}. For example, if each variable is binary ($x_i = \pm 1$) and we measure all of the averages $\langle x_i\rangle_\text{exp}$ and pairwise correlations $\langle x_ix_j\rangle_\text{exp}$, then we have the pairwise maximum entropy model,
\begin{equation}
\label{eq_pair}
P_\text{pair}(\bm{x}) = \frac{1}{Z}\text{exp}\bigg[\sum_i h_i x_i + \sum_{ij} J_{ij} x_ix_j\bigg],
\end{equation}
which is equivalent to an Ising model with all-to-all interactions \cite{brush1967history}. Importantly, since all other sources of order are explicitly removed, any predictions that the model makes must arise exclusively from the features that the modeler chose to include. In this way, models like Eq.~(\ref{eq_pair}) allow us to determine which of a system's collective features can be understood as emerging from simple correlations. This approach has provided effective quantitative descriptions for a wide range of collective phenomena in biology, including patterns of activity in neural networks \cite{Schneidman-01, Ashourvan-01, Rosch-01, Meshulam-03, Tkacik-02, Marre-01}, sequence-structure maps in proteins \cite{Marks-01, Weigt-01, Russ-01, Morcos2011}, expression patterns in genetic networks \cite{Lezon-01, Dixit-01}, folding in chromatin \cite{DiPierro-01, Lin-03, Shi-01, Messelink-01, Farre-01}, dynamics in flocks of birds \cite{Bialek-01, Cavagna-01, Bialek-02}, and activity in social networks \cite{Lynn-04}. However, we still do not know which set of features---and therefore which model---provides the best description of a system.

\subsection{Minimizing entropy}

Each choice for the set of features $\mathcal{F} = \{f_\mu(\bm{x})\}$ yields a distinct maximum entropy model $P_\mathcal{F}(\bm{x})$. Given constraints on the allowed sets of features---for example, on the number of features or their functional form---we would like to identify the optimal set $\mathcal{F}^*$ that provides the best compression of the system $P_{\mathcal{F}^*}(\bm{x})$. Due to the Boltzmann form of Eq.~(\ref{eq_PF}), the description length of a maximum entropy model can be expressed as
\begin{equation}
L(P_\mathcal{F}) = -\langle \log P_\mathcal{F}(\bm{x})\rangle_\chi = \log Z - \sum_{\mu = 1}^M \lambda_\mu \langle f_\mu(\bm{x})\rangle_\chi.
\end{equation}
Moreover, since the datasets in $\chi$ are constrained to match the experimental averages of the features in $\mathcal{F}$, we have $\langle f_\mu(\bm{x})\rangle_\chi = \langle f_\mu(\bm{x}) \rangle_\text{exp}$. Finally, due to the maximum entropy constraints in Eq.~(\ref{eq_f}), we know that the model also matches the same averages. Together, these results reveal that the description length of any maximum entropy model is equivalent to the entropy of the model itself,
\begin{align}
\label{eq_LP1}
L(P_\mathcal{F}) &= \log Z - \sum_{\mu = 1}^M \lambda_\mu \langle f_\mu(\bm{x})\rangle \\
\label{eq_LP2}
&= -\langle \log P_\mathcal{F}(\bm{x}) \rangle = S(P_\mathcal{F}).
\end{align}

We have arrived at a strikingly simple picture: minimizing the description length $L(P_\mathcal{F})$ is equivalent to minimizing the entropy $S(P_\mathcal{F})$. Thus, among all of the allowed sets of features $\mathcal{F}$, the optimal set $\mathcal{F}^*$ is the one that produces the maximum entropy model $P_{\mathcal{F}^*}(\bm{x})$ with minimum entropy $S(P_{\mathcal{F}^*})$. This is the \textit{minimax entropy principle}, which can be stated mathematically,
\begin{align}
\label{eq_F*1}
\mathcal{F}^* &= \arg\min_\mathcal{F} S(P_\mathcal{F}) \\
\label{eq_F*2}
&= \arg\min_\mathcal{F} \big\{\max_P S(P) \\
&\quad\quad\quad\quad\quad\quad :\, \langle f_\mu (\bm{x})\rangle = \langle f_\mu (\bm{x})\rangle_\text{exp} \,\, \forall\,\, f_\mu \in \mathcal{F}  \big\}. \nonumber
\end{align}
This establishes a unifying perspective: by minimizing the length of our encoding, one can derive not only the optimal model for a given set of features (maximum entropy), but also the optimal features themselves (minimax entropy).

\subsection{General solution}

The form of the minimax entropy problem in Eqs.~(\ref{eq_F*1}-\ref{eq_F*2}) immediately hints at an algorithmic solution:
\begin{enumerate}
\item For a given set of features $\mathcal{F} = \{f_\mu(\bm{x})\}$, one must first solve the traditional maximum entropy problem for the parameters $\lambda_\mu$ such that the model $P_\mathcal{F}(\bm{x})$ matches the specified features [Eq.~(\ref{eq_f})].
\item One must then repeat step 1 for each allowed set of features $\mathcal{F}$ in order to identify the optimal set $\mathcal{F}^*$ that minimizes the model entropy $S(P_\mathcal{F})$.
\end{enumerate}
This algorithm, while prohibitively inefficient in most cases (see below), provides a general solution to the minimax entropy problem.

To illustrate this solution, consider a three-spin Ising model with collective states $\bm{x} = \{x_1,x_2,x_3\}$, where $x_i = \pm 1$, defined by the Boltzmann distribution
\begin{equation}
\label{eq_Ising}
P_\text{true}(\bm{x}) = \frac{1}{Z}\exp (3x_1x_2 - 2x_2x_3 + x_1x_3).
\end{equation}
Without any knowledge of the system, the maximum entropy distribution $P(\bm{x})$ is uniform over the eight states $\bm{x}$, with entropy $S(P) = 3$ bits. This model of independent spins captures none of the underlying structure. By contrast, if we measure all three of the pairwise correlations $\langle x_ix_j\rangle_\text{true}$, then the maximum entropy model is exact $P(\bm{x}) = P_\text{true}(\bm{x})$, with entropy $S(P) = S(P_\text{true})$.

Between these two extremes, we have a choice for which correlations $\langle x_ix_j\rangle_\text{true}$ to include in a model. To select the single correlation that provides the best compression, we construct a different maximum entropy model $P(\bm{x}) = \frac{1}{Z}\exp (J_{ij}x_ix_j)$ for each correlation, fitting the interaction $J_{ij}$ such that $\langle x_ix_j\rangle = \langle x_ix_j\rangle_\text{true}$. With partial knowledge of the system, each model yields an entropy $3~\text{bits} \ge S(P) \ge S(P_\text{true})$ [Fig.~\ref{fig:figure1}(b)]. The optimal model, with minimum entropy, is constructed from the correlation $\langle x_1x_2\rangle$ corresponding to the strongest interaction $J_{12}$. Similarly, among pairs of correlations, the optimal model identifies the two strongest interactions $J_{12}$ and $J_{23}$ [Fig.~\ref{fig:figure1}(b)]; in fact, this model nearly provides an exact description of the system, with $S(P) \approx S(P_\text{true})$. This demonstrates how, by selecting optimal features, one does not need all of the details to construct an accurate model.

\section{Optimizing accuracy and information}
\label{sec_opt}

In the previous section, we derived the minimax entropy principle as the solution to a compression problem. Here, we provide two additional perspectives \cite{zhu_minimax_1997, zhu_frame_1996, lynn2023exactly, lynn2023exact}, which converge on the same framework for selecting optimal features from data.

\subsection{Minimizing divergence}

Consider a system defined by a true underlying distribution $P_\text{true}(\bm{x})$, and suppose we have access to an experiment that provides accurate measurements of system features, such that $\langle f_\mu(\bm{x})\rangle_\text{exp} = \langle f_\mu(\bm{x})\rangle_\text{true}$. As discussed above, for any set of features $\mathcal{F} = \{f_\mu(\bm{x})\}$, the maximum entropy model $P_\mathcal{F}(\bm{x})$ in Eq.~(\ref{eq_PF}) provides the most unbiased description of the system \cite{Jaynes-01, thomas_m_cover_elements_2006}. Generally, the features $\mathcal{F}$ only provide partial information about the system, yielding a model with entropy $S(P_\mathcal{F}) \ge S(P_\text{true})$. As we increase the number of features, this maximum entropy decreases until, in the limit that $\mathcal{F}$ fully defines the system, the model is exact, and $S(P_\mathcal{F}) = S(P_\text{true})$.

\begin{figure}[t]
\centering
\includegraphics[width=\linewidth]{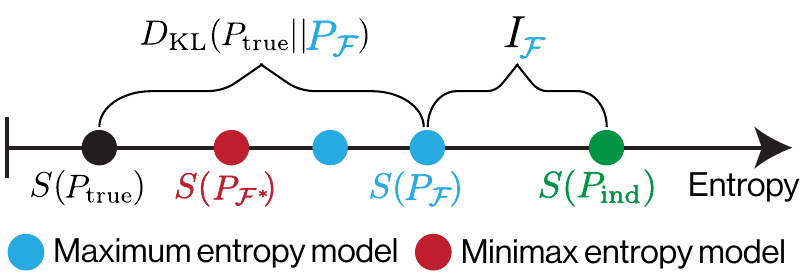}
\caption{Minimizing divergence and maximizing information. Among an allowed set of maximum entropy models (blue), the minimax entropy model has minimum entropy (red). This optimal model minimizes the KL divergence $D_\text{KL}(P_\text{true}||P_\mathcal{F}) = S(P_\mathcal{F}) - S(P_\text{true})$ with the true distribution over states (black). Beginning from an independent model (green), the optimal model also maximizes the information $I_\mathcal{F} = S(P_\text{ind}) - S(P_\mathcal{F})$ contained in correlations.}

\label{fig:optimalmaxent}
\end{figure}

These observations leave us with the intuition that the difference $S(P_\mathcal{F}) - S(P_\text{true}) \ge 0$ reflects the accuracy of the model. Indeed, this difference in entropy is precisely the Kullback-Leibler (KL) divergence between the system and the model (Fig.~\ref{fig:optimalmaxent}),
\begin{align}
\label{eq_DKL}
D_\text{KL}(P_\text{true} ||P_\mathcal{F}) &= \left< \log \frac{P_\text{true}(\bm{x})}{P_\mathcal{F}(\bm{x})} \right>_\text{true} \nonumber \\
&= \log Z - \sum_\mu \lambda_\mu \langle f_\mu(\bm{x})\rangle_\text{true} - S(P_\text{true}) \nonumber \\
&= \log Z - \sum_\mu \lambda_\mu \langle f_\mu(\bm{x})\rangle - S(P_\text{true}) \nonumber \\
&= S(P_\mathcal{F}) - S(P_\text{true}),
\end{align}
where the third equality follows from the maximum entropy constraints $\langle f_\mu(\bm{x})\rangle = \langle f_\mu(\bm{x})\rangle_\text{exp} = \langle f_\mu(\bm{x})\rangle_\text{true}$. Thus, by minimizing the entropy $S(P_\mathcal{F})$, the optimal features $\mathcal{F}^*$ also minimize the KL divergence with the true distribution (Fig.~\ref{fig:optimalmaxent}),
\begin{equation}
\mathcal{F}^* = \arg\min_\mathcal{F} D_\text{KL}(P_\text{true}||P_\mathcal{F}).
\end{equation}
In this way, the minimax entropy principle generates the most accurate description of the system $P_{\mathcal{F}^*}$ \cite{zhu_minimax_1997}.

\subsection{Maximizing information}

Each feature that we measure in experiments not only improves the accuracy of our model; it also provides information about the underlying system. For example, consider a system of binary variables $\bm{x} = \{x_1,\hdots,x_N\}$. By measuring the averages $\langle x_i\rangle_\text{exp}$, the maximum entropy model is equivalent to the independent model
\begin{equation}
P_\text{ind}(\bm{x}) = \prod_i P_i(x_i),
\end{equation}
where $P_i(x_i)$ is the marginal distribution of $x_i$. If, in addition, we measure a single correlation $\langle x_ix_j\rangle_\text{exp}$, then the maximum entropy model takes the form
\begin{equation}
P(\bm{x}) = P_{ij}(x_i,x_j)\prod_{k \neq i,j}P_k(x_k),
\end{equation}
where $P_{ij}(x_i,x_j)$ is the marginal of $x_i$ and $x_j$. Thus, by measuring $\langle x_ix_j\rangle_\text{exp}$, our uncertainty about the system decreases by an amount,
\begin{equation}
S(P_\text{ind}) - S(P) = S(P_i) + S(P_j) - S(P_{ij}),
\end{equation}
equal to the mutual information $I(x_i;x_j)$ between $x_i$ and $x_j$. If we measure all of the correlations, of all orders, then the maximum entropy model is exact. In this limit, we gain an amount of information $I_\text{tot} = S(P_\text{ind}) - S(P_\text{true})$, which is known as the total correlation or multi-information \cite{schneidman2003network}.

Between these extremes, we could measure a set of features $\mathcal{F}$ that includes the averages $\langle x_i\rangle_\text{exp}$ and a subset of correlations. Each selection of correlations provides information about the underlying system in the amount
\begin{equation}
I_\mathcal{F} = S(P_\text{ind}) - S(P_\mathcal{F}).
\end{equation}
Thus, by minimizing the entropy $S(P_\mathcal{F})$, the optimal correlations $\mathcal{F}^*$ also provide the maximum information about the system (Fig.~\ref{fig:optimalmaxent}) \cite{lynn2023exactly, lynn2023exact},
\begin{equation}
\mathcal{F}^* = \arg\max_\mathcal{F} I_\mathcal{F}.
\end{equation}
Together, these results demonstrate that the minimax entropy principle, in addition to producing optimal compressions, also yields models that are as accurate and informative as possible.

\section{Applications, solutions, and approximations}
\label{sec_review}

\subsection{Greedy algorithm}

The minimax entropy principle was originally proposed over 25 years ago in the context of machine learning \cite{zhu_minimax_1997, zhu_frame_1996}. Yet, despite decades of advances in maximum entropy modeling, the minimax entropy principle has only recently found applications in the study of complex living systems \cite{lynn2023exactly, lynn2023exact, carcamo_statistical_2024, Weaver-01}. The primary limitation is practical: While the algorithm in Sec.~\ref{sec_minimax} provides a general solution to the minimax entropy problem, a na\"{i}ve implementation is prohibitively inefficient for all but the smallest systems. For each set of features $\mathcal{F} = \{f_\mu(\bm{x})\}$, our ability to calculate the maximum entropy parameters $\lambda_\mu$ is fundamentally limited by our ability to compute model averages $\langle f_\mu(\bm{x})\rangle$. Without simplifying assumptions, this often requires computationally intensive Monte Carlo simulations \cite{nguyen2017inverse, Sethna-01, brush1967history, ferrenberg1991critical}. Then, we must repeat this process for every allowed set of features $\mathcal{F}$; however, this number can explode combinatorially. For example, if our goal is to find the $M$ optimal features out of $M' > M$ possibilities, then the number of allowed sets $\mathcal{F}$ is ${M'}\choose{M}$, which grows super-exponentially with $M$. Thus, a brute-force search over sets of features $\mathcal{F}$ is often infeasible \cite{parker2014discrete}.

Instead, we propose a greedy algorithm that grows the set of features by iteratively selecting the optimal feature at each step:
\begin{enumerate}
\item To begin, select the single allowed feature $f_1(\bm{x})$ whose empirical average $\langle f_1(\bm{x})\rangle_\text{exp}$ yields the maximum entropy model with minimum entropy.
\item After selecting $m < M$ features $f_1(\bm{x}), \hdots, f_m(\bm{x})$, search over the remaining allowed features to find the one $f_{m+1}(\bm{x})$ that yields the lowest entropy $S(P_\mathcal{F})$, where $\mathcal{F} = \{f_1(\bm{x}), \hdots, f_m(\bm{x}), f_{m+1}(\bm{x})\}$.
\item After $M$ steps, we have converged on the locally optimal set of features $\mathcal{F}^* = \{f_1(\bm{x}),\hdots,f_M(\bm{x})\}$.
\end{enumerate}
Although this greedy algorithm is not generally guaranteed to converge to the globally optimal features \cite{parker2014discrete, jungnickel2005graphs}, it provides an approximate solution with a significant increase in efficiency \cite{carcamo_statistical_2024, zhu_frame_1996, zhu_minimax_1997}. In turn, this opens the door for practical applications of the minimax entropy principle.

\subsection{Minimax entropy Gaussians}

Perhaps the simplest application of the minimax entropy principle is to systems of $N$ continuous variables $\bm{x} = \{x_1, \dots, x_N \}$ with experimental covariances $\Sigma_{ij} = \langle x_ix_j\rangle_\text{exp} - \langle x_i\rangle_\text{exp}\langle x_j\rangle_\text{exp}$. A subset of these covariances can be visualized as a network $G$ with nodes representing the different variables and edges defining the covariances that we include in our model. Given a network of covariances $G$, the maximum entropy model takes the form of a Gaussian,
\begin{equation}
P_G(\bm{x}) = \sqrt{\frac{|J|}{(2\pi)^N}} \exp \bigg[ -\sum_{(ij) \in G} J_{ij}x_ix_j \bigg],
\end{equation}
where the sum runs over all of the edges in $G$, and $|\cdot|$ is the determinant. Since the entropy of a Gaussian does not depend on the means of the variables, these can be subtracted from the data without loss of generality. If we include all of the covariances in our model, yielding an all-to-all network $G$, then the maximum entropy parameters are given directly by $J = \Sigma^{-1}$. More generally, for each covariance that we include in the model (that is, for each edge $(ij)$ in $G$), the parameter $J_{ij}$ must be calculated so that the model covariance $(J^{-1})_{ij}$ matches the measured value $\Sigma_{ij}$. For covariances not constrained in the model, $J_{ij} = 0$. Each subset of covariances (equivalently, each network $G$) therefore yields a Gaussian graphical model, which have found widespread applications across biology, psychology, and economics \cite{edwards2012introduction, yuan2007model, epskamp2018gaussian, wang2016fastggm, Uhler2017}.

\begin{figure}
    \centering
    \includegraphics[width=\linewidth]{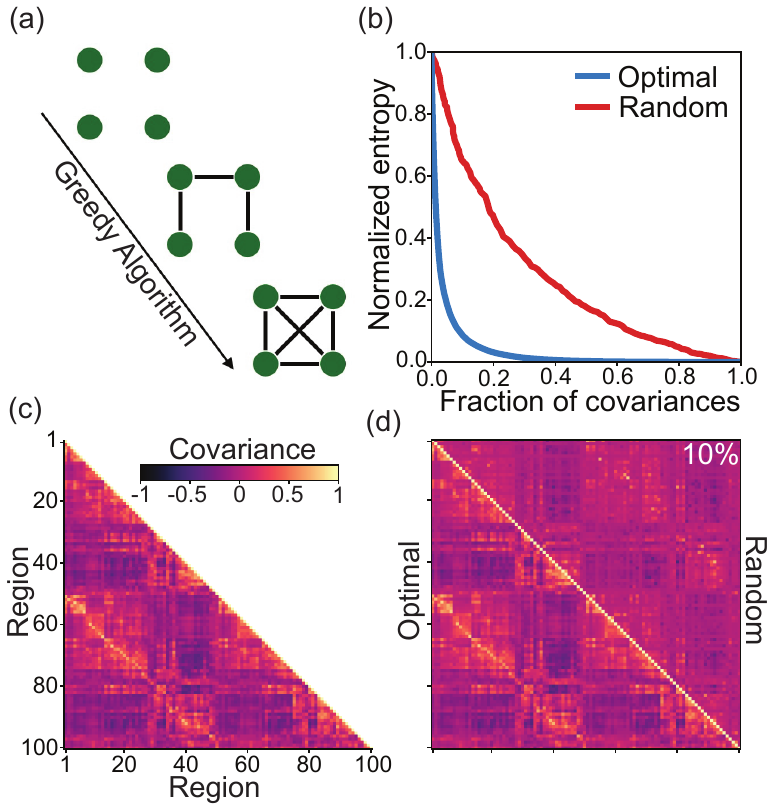}
    \caption{Minimax entropy models of human neural activity. (a) To select an optimal set of covariances, the greedy algorithm iteratively adds edges to a network $G$ so as to minimize the entropy $S(P_G)$. (b) Normalized entropy $(S(P_G) - S(P_\text{tot})/(S(P_\text{ind}) - S(P_\text{tot}))$, where $P_\text{tot}$ is the full Gaussian consistent with all of the covariances $\Sigma$, versus the number of covariances included in the model for networks constructed using the greedy algorithm (blue) and for random networks (red). Neural activity is recorded using fMRI from the cortex of 100 healthy human subjects as part of the Human Connectome Project \cite{van2013wu}. (c) Measured covariances between brain regions $\Sigma$. (d) Covariances predicted by models constructed from the optimal network (left) and a random network (right), each containing 10\% of the experimental covariances. Adapted from \cite{Weaver-01}.}
    \label{fig:Gaussian}
\end{figure}

But given covariances $\Sigma$ measured in data, which should we include in a model? The minimax entropy principle tells us that we should select the network $G$ that yields the model with the lowest entropy
\begin{equation}
S(P_G) = -\frac{1}{2}\log |J| + \frac{N}{2}\log (2\pi e),
\end{equation}
or, equivalently, the largest determinant $|J|$. Given a desired number of covariances $M$, we can use the greedy algorithm described above to grow a locally optimal network $G$ with $M$ edges [Fig.~\ref{fig:Gaussian}(a)]. In fact, this is a common strategy for model selection in Gaussian graphical models \cite{edwards2012introduction}.

To illustrate this approach, consider neural activity in human subjects measured, for example, using functional magnetic resonance imaging (fMRI). In these recordings, the collective state of the brain is defined by a vector $\bm{x} = \{x_1,\hdots,x_N\}$, where $x_i$ represents the activity of region $i$. By studying the covariances between brain regions $\Sigma$, one can understand their functional relationships and how these relationships are altered over the course of development, between cognitive tasks, and in diseases like Schizophrenia and Alzheimer's \cite{bassett2017network, lynn_physics_2019}.

Using the minimax entropy principle, one can investigate which covariances are most important for constraining the distribution over neural states. In a large cohort of 100 subjects \cite{Weaver-01}, recent work has shown that one can achieve a large reduction in entropy with a relatively small number of covariances [Fig.~\ref{fig:Gaussian}(b)]. With only 10\% of the available covariances, the optimal model captures over 90\% of the information contained in the data, and accurately predicts most of the correlations between brain regions [Fig.~\ref{fig:Gaussian}(c-d)]. By contrast, a random selection of covariances yields a much smaller drop in entropy [Fig.~\ref{fig:Gaussian}(b)] and fails to predict many correlations in the data [Fig.~\ref{fig:Gaussian}(d)]. These results demonstrate that, by optimizing the features to include in a model, the brain may be amenable to compressed descriptions.

\subsection{Minimax entropy Ising models}

Many systems are described by discrete rather than continuous variables. For example, decreasing in scale from the whole brain to a population of neurons, within a short window of time each nerve cell $i$ is either active ($x_i = 1$) or silent ($x_i = 0$). Just as in the continuous case, a subset of the correlations $\langle x_ix_j\rangle_\text{exp}$ between pairs of neurons can be represented as a network $G$. If our set of features includes all of the average activities $\langle x_i\rangle_\text{exp}$ and a network $G$ of correlations, then the maximum entropy model takes the form
\begin{equation}
P_G(\bm{x}) = \frac{1}{Z}\exp \bigg[\sum_i h_i x_i + \sum_{(ij)\in G} J_{ij}x_ix_j \bigg].
\end{equation}
This is equivalent to an Ising model with allowed interactions defined by the network $G$. If we include all of the pairwise correlations, yielding an all-to-all network $G$, then we arrive at the pairwise model in Eq.~(\ref{eq_pair}). While this approach has provided key insights into a wide range of collective biological and physical phenomena \cite{brush1967history, Schneidman-01, Ashourvan-01, Rosch-01, Meshulam-03, Tkacik-02, Marre-01, Marks-01, Weigt-01, Russ-01, Morcos2011, Lezon-01, Dixit-01, DiPierro-01, Lin-03, Shi-01, Messelink-01, Farre-01, Bialek-01, Cavagna-01, Bialek-02, Lynn-04, presse2013principles, Mora-01}, one might hope for a more compressed description. Indeed, as the number of variables $N$ increases in large-scale experiments, the number of pairwise correlations grows as $N^2$, leading to a quadratic increase in the number of features (and thus the number of parameters) included in the model.

To compress our description of the system, the minimax entropy principle tells us that we should focus on the subset of correlations $G$ that yields the Ising model $P_G$ with the minimum entropy $S(P_G)$. However, for a given set of correlations, calculating the parameters $h_i$ and $J_{ij}$ (known as the inverse Ising problem) is notoriously difficult \cite{nguyen2017inverse}; and even if we can infer the parameters, we still need to search for the model with the lowest entropy.

\begin{figure}
    \centering
    \includegraphics[width=\linewidth]{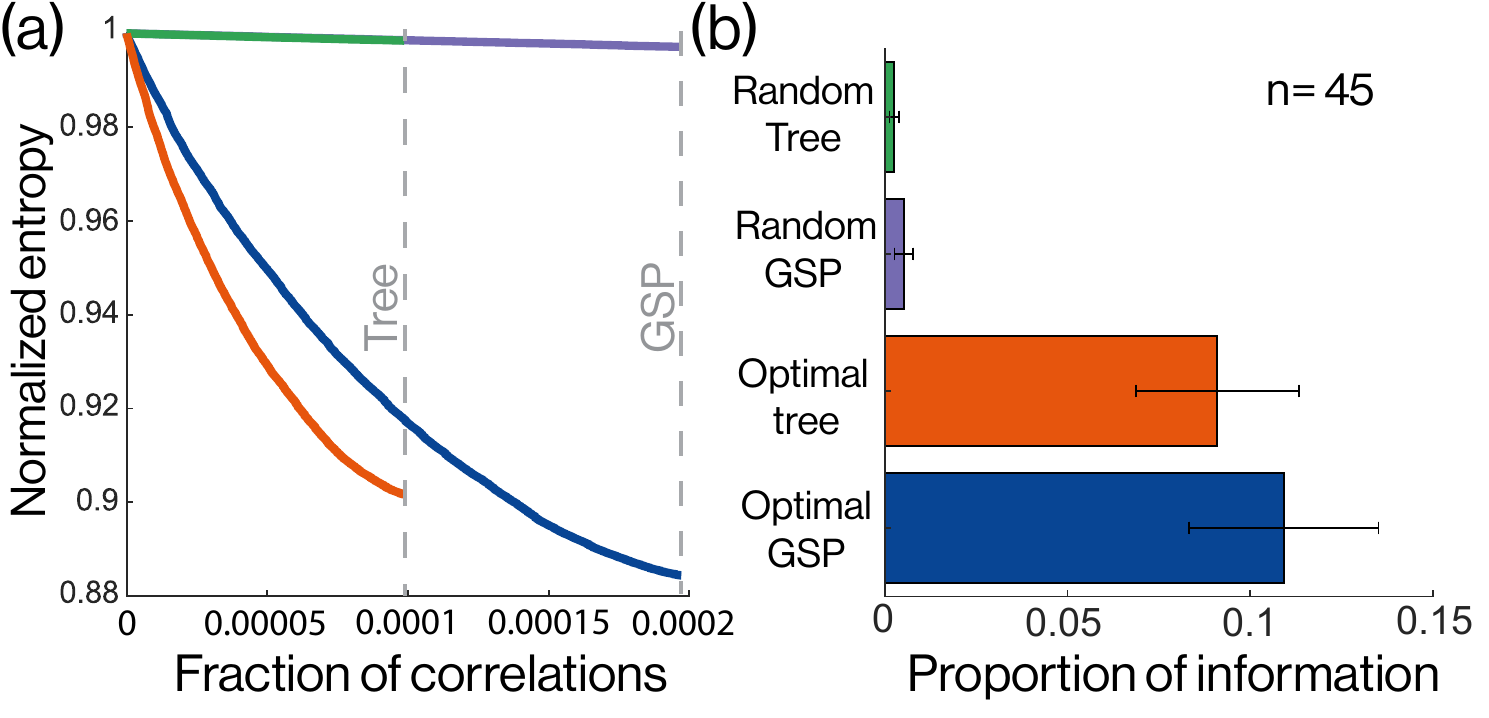}
    \caption{Minimax entropy models of large populations of neurons. (a) Entropy (normalized by the independent entropy $S(P_\text{ind})$) versus the fraction of possible correlations included in different models of collective activity in a population of $N \approx 10,000$ neurons in the mouse visual cortex \cite{stringer_high-dimensional_2019, carcamo_statistical_2024}. Using the greedy algorithm, we select the optimal tree of correlations (orange) and the optimal GSP network (blue). We compare against a random tree (green) and a random GSP network (purple). Dashed lines indicate the maximum number of correlations allowed in trees and GSP networks. (b) Information $I_G$ (normalized by $S(P_\text{ind})$) captured by different networks of correlations. Values and error bars represent averages and standard deviations across 45 different recordings \cite{stringer_high-dimensional_2019, carcamo_statistical_2024}.}
    \label{fig:tree}
\end{figure}

Calculations in statistical physics are difficult in part due to feedback loops in the interactions $J_{ij}$; eliminate these loops and calculations often become tractable \cite{baxter2016exactly}. In fact, by focusing on networks $G$ without loops (known as trees), the information contained in a set of correlations can be decomposed into a sum of mutual information \cite{chow1968approximating}
\begin{equation}
I_G = S(P_\text{ind}) - S(P_G) = \sum_{(ij)\in G} I(x_i;x_j).
\end{equation}
This decomposition reveals that the optimal tree (which minimizes $S(P_G)$ and maximizes $I_G$) is the one with the largest total mutual information. Finding this optimal network is a minimum spanning tree problem, which admits a number of efficient solutions \cite{moore2011nature}, including the greedy algorithm described previously. Moreover, once the optimal tree is identified, one can analytically calculate the corresponding parameters $h_i$ and $J_{ij}$, thus solving the inverse Ising problem \cite{nguyen2017inverse, chow1968approximating}. Together, these results present an exact and efficient solution to the minimax entropy problem, enabling optimal Ising models of very high-dimensional systems \cite{lynn2023exactly, lynn2023exact}.

The above solution hinges on a method known as exact renormalization, which is typically applied to systems without loops, such as one-dimensional Ising models or Bethe lattices \cite{rosten2012fundamentals, baxter2016exactly}. Yet this method can be extended to a broader class of networks, known as generalized series-parallel (GSP) networks, which contain loops of interactions \cite{carcamo_statistical_2024}. In fact, this is the maximal class of Ising models that can be solved using exact renormalization. With GSP networks, one can include $\sim 2N$ correlations from data, twice as many as a tree but still a vanishingly small fraction of the $\sim N^2/2$ possible correlations. As $N$ increases in high-dimensional data, can such sparse models capture anything meaningful about a system?

To answer this question, we study 45 recordings of $\sim 10,000$ neurons in the mouse visual cortex \cite{stringer_high-dimensional_2019, stringer_spontaneous_2019}. In Fig.~\ref{fig:tree}, we compare the minimax entropy models constructed from optimal trees and optimal GSP networks of correlations with maximum entropy models constructed from random trees and random GSP networks. When the optimal tree and GSP models are saturated with edges, they contain $0.02\%$ and $0.04\%$ of the available pairwise correlations, respectively. Despite this sparsity, by focusing on the most important correlations, these optimal models capture $9\%$ and $11\%$ of the independent entropy $S(P_\text{ind})$ in the populations [Fig.~\ref{fig:tree}(b)]. For comparison, random models with the same numbers of correlations provide almost no benefit over independent models. Together, these results demonstrate that large amounts of information can be packed into sparse networks of correlations, providing hope for highly compressed descriptions of neural activity.

\subsection{Minimax entropy texture modeling}

The minimax entropy principle was first developed for texture modeling in computer vision \cite{zhu_frame_1996, zhu_minimax_1997}. The task of modeling texture is to construct a distribution over images that is constrained to some of their characteristic features, which are defined by filters. For example, some filters detect vertical edges, while others detect horizontal edges. Given a set of filters, the maximum entropy principle provides the most unbiased distribution over images; yet the modeler is still tasked with choosing the set of filters. The minimax entropy principle reveals that the optimal filters, which provide the most accurate distribution over images, are the ones that produce the maximum entropy model with minimum entropy \cite{zhu_frame_1996, zhu_minimax_1997}. 
\begin{figure}[t!]
    \centering
    \includegraphics[width=\linewidth]{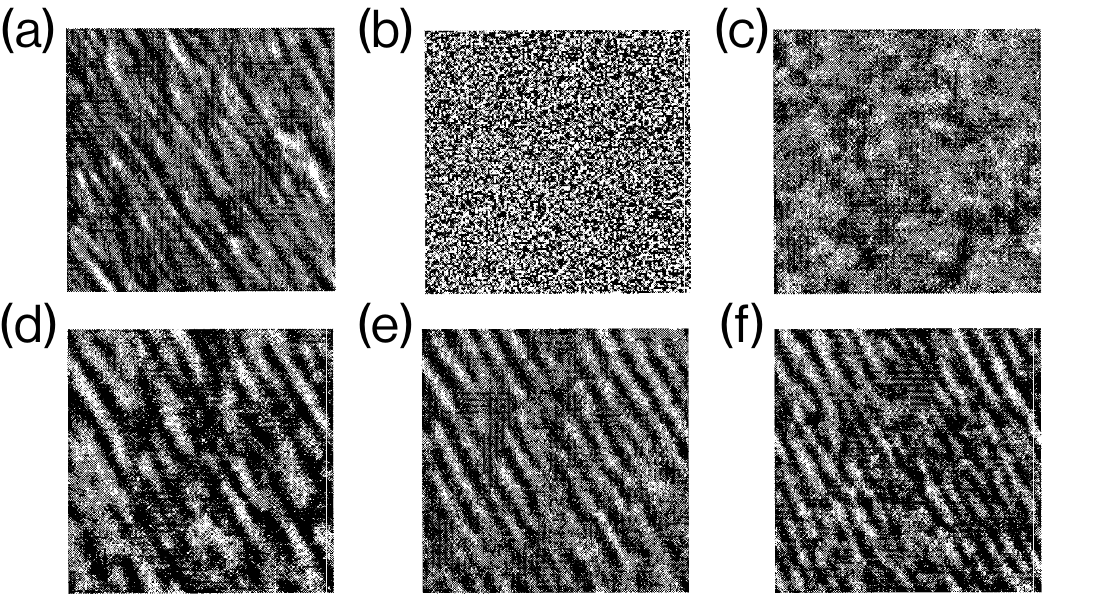}
    \caption{Minimax entropy model of texture. (a) Target image of fur. (b-e) Images generated from models with zero (b), one (c), two (d), three (e), and six (f) filters. Adapted from \cite{zhu_frame_1996}.} % Note sure if this is the right citation
    \label{fig:minimaxiamage}
\end{figure}

This method is illustrated in Fig.~\ref{fig:minimaxiamage} for an image of fur [Fig.~\ref{fig:minimaxiamage}(a)]. With no filters, the model contains no information, and we are left with an image of stochastic noise [Fig.~\ref{fig:minimaxiamage}(b)]. Using the greedy algorithm described above, one proceeds to iteratively select filters that produce the largest reduction in entropy. As more filters are added, the original image quickly comes into focus [Fig.~\ref{fig:minimaxiamage}(c-e)], until, with only six filters, the model accurately reproduces the desired texture [Fig.~\ref{fig:minimaxiamage}(f)]. Thus, just as in the neural systems described above, only a small subset of features are needed to capture large amounts of variability in the data, thereby providing a good compression.

\subsection{Minimax entropy for multiple distributions}

While the minimax entropy principle provides a systematic approach to identify the optimal subset of features from a larger set, the modeler is still tasked with defining this larger set of allowed features. In full generality, one could allow any function $f(\bm{x})$ over states. Given this flexibility, the minimum entropy is achieved by selecting $f(\bm{x})$ so that the model matches the empirical distribution over states, $P(\bm{x}) = P_\text{exp}(\bm{x})$. However, in high-dimensional data, this empirical distribution can be vastly undersampled; and even if the states are well-sampled, simply matching the empirical distribution does not provide a compressed description of the system.

Here, we consider a generalization of the minimax entropy principle, where the modeler seeks to select a set of features that apply across multiple distributions. This approach is most relevant for modeling normalized abundance counts, which arise across biology and engineering, including gene expression, microbial ecology, and image pixel intensities \cite{Dixit-01, David_Host_Genome, Rowley_Neural_1998, Zhang_Understanding_2010, deng_mnist_2012}. For concreteness, we consider $A$ different models $P_\alpha(\bm{x})$, where $\alpha = 1,\hdots, A$, each sharing the same $M$ features $f_\mu(\bm{x})$, where $\mu = 1,\hdots, M$. For a given choice of features, each model is defined by a maximum entropy distribution \cite{dixit_thermodynamic_2020}
\begin{equation}
P_\alpha(\bm{x}) = \frac{1}{Z_\alpha} \exp \bigg[ \sum_{\mu = 1}^M \lambda_{\alpha \mu} f_\mu (\bm{x})\bigg].
\end{equation}
For each distribution $\alpha$, the parameters $\lambda_{\alpha \mu}$ must be calculated so that the model averages match the empirical averages,
\begin{equation}
\sum_{\bm{x}} f_\mu(\bm{x}) P_\alpha(\bm{x}) =  \langle f_\mu(\bm{x})\rangle_{\alpha, \text{exp}}\quad \text{for all } \mu.
\end{equation}

\begin{figure}[t!]
\includegraphics[width=\linewidth]{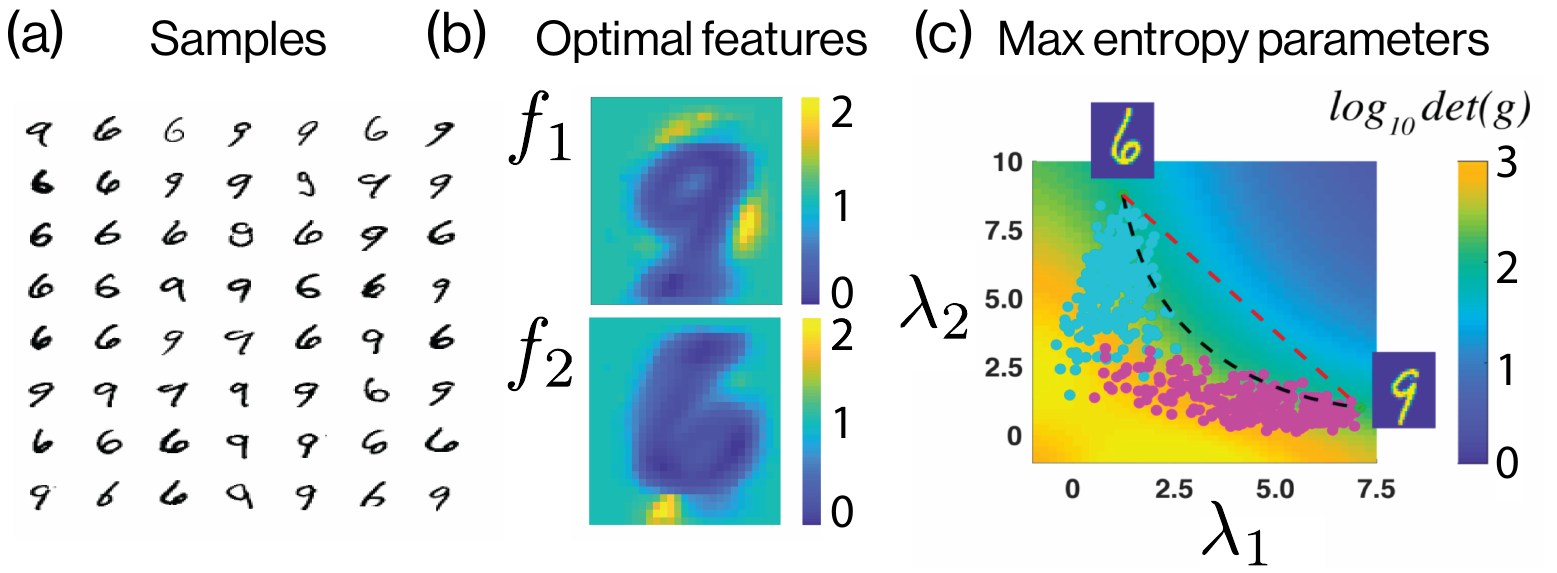}
\caption{Minimax entropy models of handwritten digits. (a) Digitized images of handwritten sixes and nines from the MNIST database \cite{deng_mnist_2012}. (b) The first two optimal features capture the salient structure in the data. (c) Each image defines a unique point in the two-dimensional space of maximum entropy parameters. Colors indicate the volume element of the Fisher-Rao metric $g$, the dashed black line illustrates the geodesic between two points in this metric, and the dashed red line represents the straight (Euclidean) line between the same two points. \label{fig_tmi}}
\end{figure}

In Fig.~\ref{fig_tmi}, we illustrate how optimal features can be inferred for multiple distributions using digitized images of handwritten sixes and nines from the MNIST database \cite{deng_mnist_2012}. Each image in Fig.~\ref{fig_tmi}(a) is defined by normalized pixel intensities, which can be treated as an empirical histogram. Using the minimax entropy principle, the first two optimal features $f_\mu(\bm{x})$ clearly reflect the underlying structure in the data [Fig.~\ref{fig_tmi}(b)]. For each image $\alpha$, we also infer the maximum entropy parameters $\lambda_{\alpha\mu}$, thus defining a point in a two-dimensional space spanning the two features [Fig.~\ref{fig_tmi}(c)]. Once again, this example demonstrates how a large amount of information can be compressed into a small number of features.

\section{Future directions and open challenges}
\label{sec_future}

\subsection{Applications}

Maximum entropy models have become increasingly applied as a principled way to map experimental data to quantitative predictions across biology, neuroscience, physics, and engineering \cite{Jaynes-01, Schneidman-01, Ashourvan-01, Rosch-01, Meshulam-03, Tkacik-02, Marre-01, Marks-01, Weigt-01, Russ-01, Morcos2011, Lezon-01, Dixit-01, DiPierro-01, Lin-03, Shi-01, Messelink-01, Farre-01, Bialek-01, Cavagna-01, Bialek-02, presse2013principles, Mora-01, phillips2004maximum, ratnaparkhi1999learning}. In each of these contexts, the minimax entropy principle has the potential to provide immediate improvements by identifying the optimal set of features---and therefore the optimal model---to describe the underlying system.

For example, gene regulatory networks are often inferred indirectly from correlations in gene expressions \cite{Lezon-01}. Rather than including all of the correlations or selecting networks based on known interactions, one can use the minimax entropy principle as a data-based method for identifying the correlations---and therefore the underlying interactions---that provide the most accurate predictions for gene co-expression. Similarly, DNA folds into compact structures known as chromatin, yielding correlations in the physical locations of genetic sites; yet the mechanisms are not fully understood. While maximum entropy approaches have been used to infer all-to-all interactions \cite{DiPierro-01, Lin-03, Shi-01, Messelink-01}, one might suspect that the underlying network is sparse. To test this hypothesis, one needs to infer the optimal network of interactions for a given level of sparsity. In precisely this way, the minimax entropy principle provides the tools to infer compressed descriptions and sparse networks of interactions not only in gene expression and chromatin structure, but also in the brain \cite{Schneidman-01, Ashourvan-01, Rosch-01, Meshulam-03, Tkacik-02, Marre-01}, animal populations \cite{Bialek-01, Cavagna-01, Bialek-02}, social networks \cite{Lynn-04}, proteins \cite{Marks-01, Weigt-01, Russ-01, Morcos2011}, and other complex living systems.

\subsection{Challenges}

The primary challenge in applying the minimax entropy principle is practical: For each set of features $\mathcal{F} = \{f_\mu(\bm{x})\}$, one must compute the parameters $\lambda_\mu$ to define the maximum entropy model $P_\mathcal{F}(\bm{x})$; then, one must repeat this process many times to search for the optimal features $\mathcal{F}^*$ that minimize the entropy $S(P_\mathcal{F})$. To solve the maximum entropy problem for the parameters $\lambda_\mu$, one must first be able calculate the model averages $\langle f_\mu(\bm{x})\rangle$. For a Gaussian model, computing covariances amounts to inverting the interaction (or precision) matrix, so the maximum entropy problem can be solved efficiently \cite{edwards2012introduction, yuan2007model}. But for systems with discrete states, such as the Ising model, computing model averages is generally difficult, requiring computationally intensive Monte Carlo simulations \cite{nguyen2017inverse, Sethna-01, brush1967history, ferrenberg1991critical}. As discussed above, one solution is to focus on networks of interactions that allow for exact solutions \cite{baxter2016exactly, lynn2023exact, lynn2023exactly}; yet these methods can impose severe restrictions on network topology. Another potential solution lies in mean-field approximations, which replace stochastic averages with deterministic self-consistency equations \cite{Sethna-01, thouless1977solution, tanaka1998mean, roudi2009ising}. However, increasing evidence suggests that some biological systems may be poised near criticality \cite{beggs2003neuronal, tkacik_thermodynamics_2015, lynn2023exact, Bialek-02, Meshulam-01}, which can lead to catastrophic failures in mean-field approximations \cite{di2025extended}. Future work is therefore needed to develop efficient techniques for solving the maximum entropy problem \cite{nguyen2017inverse, cocco_adaptive_2012}, particularly in high-dimensional systems and parameter regimes that are relevant for biology.

Even if we can compute the model $P_\mathcal{F}(\bm{x})$ for a given set of features $\mathcal{F}$, one still needs to search for the optimal set of features $\mathcal{F}^*$. As discussed previously, searching for the optimal features by brute force is intractable for all but the smallest systems \cite{parker2014discrete}. Instead, one typically relies on a greedy algorithm, which converges to a locally optimal solution, but is not guaranteed to find the global optimum \cite{jungnickel2005graphs, carcamo_statistical_2024, zhu_frame_1996, zhu_minimax_1997}. In some cases, however, it is possible to derive performance guarantees. For example, for trees of pairwise correlations, the minimax entropy problem can be reduced to a minimum spanning tree problem, and the greedy algorithm identifies the globally optimal correlations (Sec.~\ref{sec_review}). More generally, if including more features in the set $\mathcal{F}$ provides decreasing reductions in entropy $S(P_\mathcal{F})$, then the optimization problem is known as submodular \cite{fujishige2005submodular}. For submodular functions (the discrete analogue of convex functions), the greedy algorithm is guaranteed to converge to a solution that is within $1 - 1/e$ of the global optimum \cite{krause2014submodular}. Understanding when the minimax entropy problem is submodular---and thus when the greedy algorithm is near optimal---remains a clear open challenge.

\subsection{Parameterized features}

Thus far, we have focused on selecting among discrete features $\langle f_\mu(\bm{x})\rangle_\text{exp}$, such as average states $\langle x_i\rangle_\text{exp}$ or pairwise correlations $\langle x_ix_j\rangle_\text{exp}$. Instead, one could imagine optimizing over a parameterized family of functions $f_{\bm{w}}(\bm{x})$, where $\bm{w}$ is a vector of parameters that can be tuned continuously. For example, consider the family of features $f_{\bm{w}}(\bm{x}) = \sum_i w_ix_i$, which yields the maximum entropy model
\begin{equation}
P_{\bm{w}}(\bm{x}) = \frac{1}{Z}\exp\bigg[\lambda\sum_i w_ix_i\bigg],
\end{equation}
where $\lambda$ must be calculated so that
\begin{equation}
\sum_i w_i \langle x_i\rangle = \sum_i w_i \langle x_i\rangle_\text{exp}.
\end{equation}
Now, rather than selecting between discrete features, we want to find the values for $\bm{w}$ that produce the maximum entropy model with minimum entropy $S(P_{\bm{w}})$. For binary variables, one can show that these optimal parameters $\bm{w}^*$ are the ones that give the correct independent model, such that $P_{\bm{w}^*}(\bm{x}) = P_\text{ind}(\bm{x})$. 

More generally, by focusing on parameterized features, one can take advantage of methods from continuous optimization. For example, one can construct a gradient descent algorithm by computing the the gradients $\nabla_{\bm{w}} S(P_{\bm{w}})$. This continuous version of the minimax entropy problem may provide complementary insights to the original discrete version, which have yet to be explored.

\section{Conclusions}
\label{sec_concl}

When constructing models, there exists an inherent trade-off between accuracy and simplicity. A good model is a good compression, providing an accurate description with as few details as possible. From an information-theoretic perspective, accurate models define efficient encodings of data, thus minimizing description length. Given a set of details---or features measured in experiments---the description length is minimized by the model with maximum entropy \cite{feder_maximum_1986}. While this has long been understood, it does not tell us which features we should choose to begin with.

Here we show that the optimal features, which minimize description length, are the ones that yield the maximum entropy model with minimum entropy. Thus, for a given number of features, the minimax entropy principle provides the optimal compression. 

Although the minimax entropy principle was proposed decades ago \cite{zhu_frame_1996, zhu_minimax_1997}, it has only recently been applied in the study of complex living systems \cite{lynn2023exactly, lynn2023exact, carcamo_statistical_2024, Weaver-01}. Indeed, applications are limited by our ability to solve statistical physics problems and then search over many solutions. These practical challenges become even more daunting as experiments grow to record from larger systems. Future success therefore depends on the alignment of advanced techniques from theoretical physics with their appropriate contexts in biology, neuroscience, and engineering.

\begin{acknowledgments}
We thank W.~Bialek, S.E.~Palmer, F.~Mignacco, L.~Di Carlo, Q.~Yu, M.P.~Leighton, B.~Machta, and C.~Smith for helpful discussions. P.D. acknowledges support from the National Institutes of Health (NIH/NIGMS R35GM142547). 
\end{acknowledgments}

\appendix

\section{Maximum entropy from minimum description length}
\label{app_MDL}

Consider a system with states $\bm{x}$ and a set of $M$ functions $\{f_\mu(\bm{x})\}$, where $\mu=1, \hdots,M$. From experiments, we have access to the empirical averages $\langle f_\mu (\bm{x}) \rangle_\text{exp}$. These features could have been generated by one of many datasets $\{\bm{x}^{(t)}\}$ consisting of $T$ samples of the system state $\bm{x}^{(t)}$, where $t = 1,\hdots, T$. The set of all datasets consistent with the observed features is denoted $\chi$ [Eq.~(\ref{eq_chi})].

For every state $\bm{x}$, a model $P(\bm{x})$ defines a code word of length $\ell_P(\bm{x}) = -\log P(\bm{x})$ \cite{Shannon-01, thomas_m_cover_elements_2006}. By averaging over all of the possible datasets in $\chi$, we arrive at the description length of the model $L(P)$ [Eq.~(\ref{eq_L})]. To compute this average, we note that each dataset $\{\bm{x}^{(t)}\}$ in $\chi$ defines a histogram
\begin{equation}
Q(\bm{x}) = \frac{1}{T} \sum_{t = 1}^T \delta_{\bm{x}, \bm{x}^{(t)}}.
\end{equation}
The description length can therefore be written
\begin{equation}
L(P) = \langle \ell_P(\bm{x})\rangle_\chi = \frac{1}{|\chi|}\sum_{\{\bm{x}^{(t)}\} \in \chi} \langle \ell_P(\bm{x})\rangle_Q.
\end{equation}
For each dataset, $TQ(\bm{x})$ counts the number of appearances of the state $\bm{x}$. Thus, $\frac{T!}{\prod_{\bm{x}} (TQ(\bm{x}))!}$ counts the number of datasets $\{\bm{x}^{(t)}\}$ corresponding to a given histogram $Q(\bm{x})$. We can therefore rewrite the average over datasets as a weighted average over histograms,
\begin{equation}
\label{eq_LQ}
L(P) = \frac{1}{|\chi|}\sum_{Q\in \mathcal{Q}} \frac{T!}{\prod_{\bm{x}} (TQ(\bm{x}))!} \langle \ell_P(\bm{x})\rangle_Q,
\end{equation}
where $\mathcal{Q}$ is the set of histograms generated by datasets in $\chi$.

In the limit of large $T$, one can use Stirling's approximation to show that
\begin{equation}
\frac{T!}{\prod_{\bm{x}} (TQ(\bm{x}))!} \approx e^{TS(Q)},
\end{equation}
where $S(Q)$ is the entropy of $Q(\bm{x})$ [Eq.~(\ref{eq:ent})]. Thus, in the same limit, the sum in Eq.~(\ref{eq_LQ}) becomes dominated by the histogram $Q^*(\bm{x})$ with the largest entropy, such that
\begin{equation}
L(P) \approx \langle \ell_P(\bm{x}) \rangle_{Q^*}.
\end{equation}
We recognize this as the cross entropy between the maximum entropy distribution $Q^*(\bm{x})$ and the model $P(\bm{x})$, which can be decomposed into two terms
\begin{equation}
L(P) = S(Q^*) + D_\text{KL}(Q^*||P).
\end{equation}
Thus, among all of the distributions $P(\bm{x})$ consistent with the measured features, the one that minimizes the description length $L(P)$ is the one with maximum entropy $Q^*(\bm{x})$ \cite{feder_maximum_1986}.

\section{Maximum entropy distribution}
\label{app_maxEnt}

Given a set of $M$ features $\mathcal{F} = \{f_\mu(\bm{x})\}$, the maximum entropy model $P_\mathcal{F}(\bm{x})$ is defined by the Boltzmann distribution \cite{Jaynes-01, thomas_m_cover_elements_2006}. To derive this functional form, one begins with the Lagrangian
\begin{align}
\mathcal{L}(P) = S(P) &+ \sum_{\mu = 1}^M \lambda_\mu \bigg( \left<f_\mu(\bm{x}) \right>_\text{exp} - \sum_{\bm{x}} f_\mu(\bm{x}) P(\bm{x}) \bigg) \nonumber \\
&+ C\bigg(1 - \sum_{\bm{x}} P(\bm{x})\bigg),
\end{align}
where the parameters $\lambda_\mu$ are Lagrange multipliers that enforce the constraints $\left<f_\mu(\bm{x}) \right> = \left<f_\mu(\bm{x}) \right>_\text{exp}$, and $C$ ensures normalization. Setting the derivative $\frac{\partial \mathcal{L}}{\partial P(\bm{x})}$ equal to zero, one can solve for the functional form for the maximum entropy model in Eq.~(\ref{eq_PF}).

\bibliography{BibPersepctive}

\end{document}